\documentclass[journal=jctcce,manuscript=article]{achemso}
\setkeys{acs}{articletitle = true}

\usepackage[version=3]{mhchem} 
\usepackage{csvsimple}

\def\rv{{\bf r}}
\def\xv{{\bf x}}

\def\xvu{\underline{\bf x}}

\newcommand{\method}{\mathrm{FDM}}
\newcommand{\diamatA}{|\Psi^A_0|^2}
\newcommand{\diamatB}{|\Psi^B_0|^2}

\author{Derk P. Kooi}
\affiliation
{Department of Chemistry \& Pharmaceutical Sciences and Amsterdam Institute of Molecular and Life Sciences (AIMMS), Faculty of Science, Vrije Universiteit, De Boelelaan 1083, 1081HV Amsterdam, The Netherlands}
\email{d.p.kooi@vu.nl}
\author{Timo Weckman}
\affiliation
{Department of Chemistry \& Pharmaceutical Sciences and Amsterdam Institute of Molecular and Life Sciences (AIMMS), Faculty of Science, Vrije Universiteit, De Boelelaan 1083, 1081HV Amsterdam, The Netherlands}
\email{t.e.j.weckman@vu.nl}
\author{Paola Gori-Giorgi}
\affiliation
{Department of Chemistry \& Pharmaceutical Sciences and Amsterdam Institute of Molecular and Life Sciences (AIMMS), Faculty of Science, Vrije Universiteit, De Boelelaan 1083, 1081HV Amsterdam, The Netherlands}
\email{p.gorigiorgi@vu.nl}

\title[title] {Dispersion without many-body density distortion: Assessment on atoms and small molecules}

\begin{document}
\begin{abstract}
We have implemented and tested the method we have recently proposed [J. Phys. Chem. Lett. 10, 1537 (2019)] to treat dispersion interactions, which is derived from a supramolecular wavefunction constrained to leave the diagonal of the many-body density matrix of each monomer unchanged. The corresponding variational optimization leads to expressions for the dispersion coefficients in terms of the ground-state pair densities of the isolated monomers only, which provides a framework to build new approximations without the need for polarizabilities or virtual orbitals. The question we want to answer here is how accurate this ``fixed diagonal matrices'' (FDM) method can be for isotropic and anisotropic $C_6$ dispersion coefficients when using monomer pair densities from different levels of theory, namely Hartree-Fock, MP2 and CCSD. For closed-shell systems, FDM with CCSD monomer pair densities yields the best results, with a mean average percent error for isotropic $C_6$ dispersion coefficients of about 7\% and a maximum absolute error within 18\%. The accuracy for anisotropic dispersion coefficients with FDM on top of CCSD ground states is found to be similar. The performance for open shell systems is less satisfactory, with CCSD pair densities not always providing the best result. In the present implementation, the computational cost on top of the monomer's ground-state calculations is $\mathcal{O}(N^4)$.

\end{abstract}

\section{Introduction}

The attractive London dispersion interaction between atoms and molecules is weaker than covalent bonding forces, but while the latter decay exponentially with the separation $R$ between the monomers, dispersion interactions decay only polynomially in $1/R$. Because of this dominating long range character, dispersion plays a crucial role in various chemical systems and processes, such as protein folding, soft solid state physics, gas--solid interfaces etc. An accurate, computationally efficient, and fully nonempirical treatment of dispersion forces remains an open challenge, and it is the objective of several ongoing efforts (see, e.g., refs~\citenum{GriHanBra-CR-16,ClaKimDixKimGouRocLeb-JCP-18,StoVooTka-CSR-19} for recent reviews and benchmarks).

We have recently introduced a class of variational wave functions that capture the long-range interactions between two quantum systems without deforming the diagonal of the many-body density matrix of each monomer.\cite{KooGor-JPCL-19}  The variational take on dispersion is certainly not new, as, for example, variational calculations of the dispersion coefficients have been performed in the context of (Hylleraas) variational perturbation theory\cite{Tha-JCP-81} and variational calculations of the dispersion energy at finite inter-monomer distance have been carried out in the framework of Symmetry-Adapted Perturbation Theory (SAPT) using orthogonal projection.\cite{KorWilBuk-JCP-97} The distinctive feature of our approach is the reduction of
dispersion to a balance between kinetic energy and monomer-monomer interactions only,  providing an explicit expression for the dispersion energy in terms of the ground-state pair densities of the isolated monomers.  

Although the supramolecular wavefunction constructed in this ``fixed diagonal matrices'' (FDM) approach can never be exact, as density distortion is prohibited, it provides a variational expression for the dispersion coefficients when accurate pair densities for the momoners are used, at a computational cost given essentially by the ground-state monomer calculations.  The FDM approach has been found to yield exact results for the dispersion coefficients up to $C_{10}$ for the H-H case (and up to $C_{30}$ for the second-order coefficients), and very accurate results (0.17\% error on $C_6$) for He-He and He-H.\cite{KooGor-JPCL-19} This is achieved by reshuffling the contributions of kinetic and potential energy inside each monomer, as shown in table 1 of ref.~\citenum{KooGor-FD-20}. Another way to look at it is the following: dispersion between two systems in their ground state is a competition between a distortion of the fragments's ground-state (which raises the energy with respect to $E_0^A+E_0^B$) and the interfragment interaction that can lower the energy of the two systems together. As proven by Lieb and Thirring,\cite{LieThi-PRA-86} the raise in energy due to the distortion of the fragment's ground-state can be always made quadratic with respect to a set of variational parameters, with the interfragment interaction being linear. With our FDM constraint we force the quadratic raise in energy of the isolated fragments to be of kinetic energy origin only, since only the off-diagonal elements of the monomer's reduced density matrices are allowed to change. For the special case of two ground-state one-electron  fragments, this can be showm to give the same result for the dispersion coefficients as second-order Rayleigh-Schr\"odinger perturbation theory.\cite{KooGor-FD-20}

This FDM construction is fundamentally different from approaches to incorporate dispersion based on the Adiabatic Connection (AC) and the Fluctuation Dissipation Theorem  (see, for example ref~\citenum{dispersion-book-ACFD} and references therein). In these methods the interacting system is connected to a non-interacting one with the same density via the AC formalism: the monomer's pair density changes as the electron-electron interaction is turned on. A different AC approach in which only the monomer-monomer interaction is turned on has been introduced very recently in ref.~\citenum{NguCheAge-JCTC-20}: the main difference with our FDM formalism is that in our case we keep the densities (and pair densities) of the monomers equal to their isolated ground-state value, while in ref.~\citenum{NguCheAge-JCTC-20} the density is kept equal to the one of the complex for all coupling strength values.

The FDM expressions for the dispersion coefficients in terms of the ground-state pair densities of the isolated monomers offer a neat theoretical framework to build new approximations, by using pair densities from different levels of accuracy, including exchange-correlation holes  from density functional theory. This idea is similar in spirit to the eXchange Dipole Moment (XDM) of Becke and Johnson,\cite{BecJoh-JCP-05,BecJoh-JCP-07} with the main difference that in our case we do not need the static atomic polarizabilities, as everything can be expressed in terms of ground state monomer densities and exchange-correlation holes.\cite{KooGor-FD-20}
Before considering the use of the FDM framework to build DFT-based approximations, however, one should ask the question: how accurate can this approach be if we use accurate monomer's ground-state pair densities, beyond the simple H and He cases?
The aim of this work is exactly to answer to this question by exploring the performance of the FDM expression for the dispersion $C_6$ coefficient for atoms and molecules using different levels of theory for the monomer calculations, studying the convergence and basis set dependence of the results. We test the approach on 459 pairs of atoms, ions, and small molecules, using Hartree-Fock (HF), second-order M{\o}ller-Plesset perturbation theory (MP2) and coupled cluster with singles and doubles (CCSD) ground-state pair densities. We should keep in mind that the FDM expression is guaranteed to be variational, yielding a lower bound to $C_6$, only when we use exact pair densities of the monomers. As we shall see, for closed shell systems, this is almost always the case with CCSD pair densities, which yield in general good results, slightly underestimating $C_6$, although there are exceptions. With HF pair densities, as it was already found in a preliminary result for the Ne-Ne case in ref.~\citenum{KooGor-JPCL-19}, $C_6$ is, in the vast majority of cases, overestimated. 

The paper is organised as follows. In sec~\ref{sec:theory} we illustrate our working equations, including the expressions for the isotropic $C_6$ coefficients and for the anistropies, with the computational details reported in sec~\ref{sec:compdet}. The results are discussed in sec~\ref{sec:results}, and conclusions and perspectives  in sec~\ref{sec:conc}.

\section{Theory}\label{sec:theory}
We consider two systems $A$ and $B$ separated by a (large) distance $R$ having isolated ground-state wavefunctions $\Psi_0^A(\xvu_A)$ and $\Psi_0^B(\xvu_B)$, where $\xv$ denotes the spin-spatial coordinates ($\rv, \sigma$) and $\xvu_{A/B}$ denote the whole set of the spin-spatial coordinates of electrons in system $A/B$.  The $\method$ framework is defined by the following constrained minimisation problem\cite{KooGor-JPCL-19,KooGor-FD-20}
\begin{equation}\label{eq:Levy}
	 E_{\rm disp}^\method(R)=\min_{\Psi_R\to \diamatA,\diamatB}\langle\Psi_R|\hat{T}+\hat{V}_{ee}^{AB}|\Psi_R\rangle-T_0^A-T_0^B-U[\rho_0^A,\rho_0^B],
\end{equation}
where $\hat{T}$ is the usual kinetic energy operator acting on the full set of variables $\xvu_A,\xvu_B$, and
$\hat{V}_{ee}^{AB}=\sum_{i\in A,j\in B}\frac{1}{|\rv_i-\rv_j|}$.
With $T_0^{A/B}$ we denote the ground-state kinetic energy expectation values of the two separated systems and 
\begin{equation}
U[\rho_0^A,\rho_0^B]=\int d\rv\int d\rv'\frac{\rho_0^A(\rv)\rho_0^B(\rv')}{|\rv-\rv'|},
\end{equation}
where $\rho_0^{A(B)}$ are the ground-state one-electron densities of the two systems. 
The constraint ${\Psi_R\to \diamatA,\diamatB}$ means that the search in eq~\eqref{eq:Levy} is performed over wavefunctions $\Psi_R(\xvu_A,\xvu_B)$ that leave the diagonal of the many-body density matrix of each fragment unchanged with repect to the ground-state isolated value. We work in the polarization approximation, in which the electrons in $A$ are distinguishable from those in $B$. The constrained-search formulation, eq~\eqref{eq:Levy}, makes dispersion a simple competition between kinetic energy and monomer-monomer interaction, as all the other monomer energy components cannot change by construction. This also guarantees that no electrostatic or induction contributions appear in eq~\eqref{eq:Levy}.

For the minimizer of eq~\eqref{eq:Levy} we use the variational ansatz of ref~\citenum{KooGor-JPCL-19}, 
\begin{equation}
\Psi(\xvu_A, \xvu_B) = \Psi_0^A(\xvu_A) \Psi_0^B(\xvu_B) \sqrt{1+\sum_{i\in A, j\in B} J_R(\rv_i, \rv_j)},
\end{equation}
where the function $J_R$ correlates electrons in $A$ with those in $B$, and is written in the form
\begin{equation}\label{eq:Jwithb}
J_R(\rv, \rv') = \sum_{ij} c_{ij, R} b_i^A(\rv) b_j^B(\rv'), 
\end{equation}
where $c_{ij, R}$ are parameters, which are determined variationally. The functions $b_i^{A/B}(\rv)$ for now are an arbitrary set of ``dispersal'' functions, used as basis to expand $J_R$. The constraint ${\Psi_R\to \diamatA,\diamatB}$ is enforced by imposing\cite{KooGor-JPCL-19}
\begin{align}
	& \int \rho_0^A(\mathbf{r}_{i_A})J_R(\mathbf{r}_{i_A},\mathbf{r}_{j_B}) \mathrm{d} \mathbf{r}_{i_A}= 0\qquad \forall\; \mathbf{r}_{j_B} \label{eq:densconstrA} \\
	& \int \rho_0^B(\mathbf{r}_{j_B})J_R(\mathbf{r}_{i_A},\mathbf{r}_{j_B}) \mathrm{d} \mathbf{r}_{j_B}= 0\qquad \forall\; \mathbf{r}_{i_A} \label{eq:densconstrB}
\end{align}
Thanks to this constraint, the expectation of the external potential and of the electron-electron interactions inside each monomer cancel out in the interaction energy, whose variational minimization takes a simplified form.\cite{KooGor-JPCL-19,KooGor-FD-20} If we peform the multipolar expansion of the monomer-monomer interaction, we can, accordingly, expand $c_{ij, R}$ in a series of inverse powers of $R$,
\begin{equation}\label{eq:expacij}
c_{ij, R} = c_{ij}^{(3)}R^{-3}+c_{ij}^{(4)}R^{-4}+c_{ij}^{(5)} R^{-5}+\mathcal{O}(R^{-6}),
\end{equation} 
which leads to explicit expressions for the dispersion coefficients. In this paper, we focus on the leading $C_6$ coefficient of the term $-C_6\, R^{-6}$ in the dispersion interaction energy, which is determined by the variational parameters $c_{ij}^{(3)}$ in eq~\eqref{eq:expacij}, denoted simply $c_{ij}$ in the rest of this work. 

As detailed in the supplementary material of ref~\citenum{KooGor-JPCL-19}, the variational equation for $C_6$ corresponding to our wave function is given in terms of the matrices $\tau_{ij}^{A/B}$, $S_{ij}^{A/B}$, and $P_{ij}^{A/B}$ (which determine the kinetic correlation energy),
\begin{align}
	\tau_{ij}^A & = \int \rho_0^A(\rv) \nabla b_i^A(\rv)\cdot \nabla b_j^A(\rv)\, d\rv, \label{eq:tauij} \\
	S_{ij}^A & =\int \rho_0^A(\rv) b_i^A(\rv) b_j^A(\rv)\, d\rv, \label{eq:Sij} \\
	P_{ij}^A & =  \int \mathrm{d} \mathbf{r}_{1_A} \int\mathrm{d} \mathbf{r}_{2_A} P_0^A(\mathbf{r}_{1_A}, \mathbf{r}_{2_A}) b_i^A(\mathbf{r}_{1_A}) b_j^A(\mathbf{r}_{2_A}), \label{eq:Pij}
\end{align}
with similar expressions for system $B$, and of
the matrix $w_{ij}$ (which determines the monomer-monomer interaction),
\begin{align}\label{eq:wij}
w_{ij}  = \sum_{e=x,y,z} h_e (d_{e,i}^A+D_{e,i}^A)(d_{e,j}^B +D_{e,j}^B),
\end{align}
with $e=x,y,z$, and $h_e=(1,1,-2)$ when the intermolecular axis is parallel to the $z$-axis. The vectors $\mathbf{d}_i^A$ and $\mathbf{D}_i^A$ determine the dipole--dipole interaction terms, 
\begin{align}
 \mathbf{d}_i^A & = \int \mathrm{d} \mathbf{r}_{1_A}\rho_0^A(\mathbf{r}_{1_A}) b_i^A(\mathbf{r}_{1_A})\, \mathbf{r}_{1_A} \label{eq:di}\\
 \mathbf{D}_i^A & = \int \mathrm{d} \mathbf{r}_{1_A} \int \mathrm{d} \mathbf{r}_{2_A} P_0^A(\mathbf{r}_{1_A}, \mathbf{r}_{2_A})b_i^A(\mathbf{r}_{2_A})\, \mathbf{r}_{1_A},\label{eq:Di}
\end{align}
with, again, similar expressions for monomer $B$.
In eqs~\eqref{eq:Pij} and~\eqref{eq:Di} $P_0^{A/B}$ is the ground-state pair density of the two monomers, with usual normalization to $N(N-1)$.

In our previous work\cite{KooGor-JPCL-19} the matrices $S^{A/B}_{ij} + P^{A/B}_{ij}$ were diagonalized through a L\"owdin orthogonalization among the $b_i$'s, transforming the matrices $w_{ij}$ and $\tau^{A/B}_{ij}$ accordingly.  The variational coefficients $c_{ij}$ were then determined via the solution of a Sylvester equation,\cite{BarSte-ACM-72,KooGor-JPCL-19}
\begin{equation}\label{eq:Sylv}
\sum_k \tau_{ik}^A c_{kj} + \sum_l c_{il} \tau^B_{lj}= -4\, w_{ij}.
\end{equation}
Here we diagonalize $\tau^{A/B}_{ij}$ with $S_{ij}^{A/B}+P_{ij}^{A/B}$ as a metric through a generalized eigenvalue problem, again transforming accordingly $w_{ij}$, so that the indices indicate from now on matrix elements with the transformed $b_i$'s. The advantage is that this eigenvalue problem needs to be solved only once for each monomer, while the Sylvester equation \eqref{eq:Sylv} needs to be solved for each pair $AB$.  This way we can directly obtain the variational coefficients $c_{ij}$ as
\begin{align}
\sum_k \delta_{ik}\tau_{i}^A c_{kj} + \sum_l \delta_{jl}\tau^B_{j}c_{il} &= 4 w_{ij},\\
\tau_i ^A c_{ij} +  \tau_j ^B c_{ij}  &= 4 w_{ij},\\
\Rightarrow c_{ij} &= -\frac{4 w_{ij}}{\tau^A_i + \tau^B_j}.
\end{align}
The dispersion coefficient $C_6$ then takes the simpler (and computationally faster) form
\begin{equation}\label{eq:C6ani}
C_6^{AB} =  -\sum_{ij} c_{ij} w_{ij} - \frac{1}{8} \sum_{ij} c_{ij}^2 (\tau_i^A+\tau_j^B) =  \sum_{ij}\frac{2 w_{ij}^2}{\tau^A_i + \tau^B_j}.
\end{equation}
For molecules, eq~\eqref{eq:C6ani} gives access to the orientation-dependent $C_6^{AB}$ coefficient, where the kinetic energy terms $\tau_i^{A/B}$ are clearly rotationally invariant, and the dependence on the relative orientation of the monomers enters through $w_{ij}$, as shown by eqs~\eqref{eq:wij}-\eqref{eq:Di}. In order to compare with values from the literature, it is often necessary to compute the orientation-averaged isotropic $\overline{C}_6^{AB}$ coefficients, which can be obtained by performing the orientation average directly on each $w_{ij}^2$, yielding
\begin{equation}\label{eq:C6iso}
\overline{C}_6^{AB} = \sum_{ij}  \frac{2\,\overline{w^2_{ij}}}{\tau_i^A + \tau_j^B}.
\end{equation}
The $\overline{w^2_{ij}}$ is the spherically-averaged interaction term given by
\begin{equation}
\overline{w^2_{ij}} = \frac{2}{3} \sum_{e=x,y,z} \left(d_{e,i}^A+D_{e,i}^A\right)^2 \sum_{f=x,y,z} \left(d_{f,j}^B+D_{f,j}^B\right)^2.
\end{equation}

To also assess the accuracy for the orientiation dependence, we consider the case of linear molecules, for which  one usually defines anisotropic dispersion coefficients by writing the dispersion coefficient $C_6$ as\cite{MeaKum-IJQC-90},
\begin{align}\label{eq:C6aniso}
C_6^{AB}(\theta_A,\phi_A,\theta_B,\phi_B) =& \overline{C}_6^{AB}\Big(1 + \Gamma_6^{AB} P_2(\cos(\theta_A)) + \Gamma_6^{BA}  P_2(\cos(\theta_B)) \\ &+ \Delta_6^{AB} \frac{4 \pi}{5} \sum_{m=-2}^{2} (3- |m|) Y_{2}^m(\theta_A, \phi_A) Y_2^{-m}(\theta_B, \phi_B) \Big),  \nonumber
\end{align}
where $P_n$ denotes Legendre polynomials and $Y_\ell^m$ spherical harmonics. The anisotropic dispersion coefficients  $\Gamma_6^{AB}$ and $\Delta_6^{AB}$ can be obtained from our formalism as,
\begin{align}
\Gamma_6^{AB} &= \frac{2}{3 \overline{C}_6} \sum_{ij} \frac{-\sum_{e=x,y,z} h_e (d_{e,i}^A+D_{e,i}^A)^2 \sum_{f=x,y,z}(d_{f,j}^B+D_{f,j}^B)^2}{\tau_i+\tau_j},\\
\Delta_6^{AB} &= \frac{1}{3 \overline{C}_6} \sum_{ij} \frac{\sum_{e=x,y,z} h_e (d_{e,i}^A+D_{e,i}^A)^2 \sum_{f=x,y,z} h_f (d_{f,j}^B+D_{f,j}^B)^2}{\tau_i+\tau_j}.
\end{align}
A similar expression holds for $\Gamma_6^{BA}$, but with the roles of $A$ and $B$ exchanged. 

On top of the monomer calculations, the diagonalization to compute $C_6$ scales formally as $n_A^3+n_B^3$, where $n_{A/B}$ is the number of functions $b_i^{A/B}$ needed to converge, which, however, seems so far independent of system size. We should however also mention the cost of computing the matrix elements: the most expensive part is the first step of the two-step contraction to obtain $P_{ij}$, which scales as $\mathcal{O}(N_{\rm orb}^4 n_{A/B})$, while the second step scales as $\mathcal{O}(N_{\rm orb}^2 n_{A/B}^2)$, as expensive as obtaining $S_{ij}$ and $\tau_{ij}$, where $N_{\rm orb}$ is the number of spatial orbitals used in the monomer calculations.

\section{Computational Details}\label{sec:compdet}
\subsection{Choice of the dispersal functions $b_i(\rv)$}\label{subsec:bi's}
For the dispersal functions $b_i^{A/B}(\rv)$ of eq~\eqref{eq:Jwithb} we have chosen multipoles centered in $\mathbf{r}_0 =(x_0,\, y_0,\, z_0)$,
\begin{equation}\label{eq:bi}
	b_i(\mathbf{r}) = (x-x_0)^{s_i} (y-y_0)^{t_i} (z-z_0)^{u_i}.
\end{equation}
For atoms the obvious choice for $\mathbf{r}_0$ is the position of the nucleus; for molecules, in this first exploration, we have set $\mathbf{r}_0$ at the center of nuclear mass. We include all $b_i$, such that $s_i + t_i + u_i < n_\mathrm{max}$, where $n_\mathrm{max}^{A/B}$ is a parameter, which is set equal to 22 in all our calculations, which yields in general reasonably converged results (see sec~\ref{subsec:convergence_n} for a more detailed discussion on convergence). We should remark that the choice of eq~\eqref{eq:bi} is dictated mainly by the immediate availability of integrals: our goal here is to investigate whether the method is worth or not investing in further implementation and optimization. The question on how to determine the best possible $b_i^{A/B}(\rv)$ is open, with different strategies discussed in ref~\citenum{KooGor-FD-20}. 

\subsection{Matrix elements}
We denote the spatial orbitals used in the monomer calculations by $\phi_a(\rv)$ with indices $a, b, c, d$. The spin-summed one-body reduced density matrix (1-RDM) is written as $\gamma_{ab}$,
\begin{equation}
\gamma(\mathbf{r}, \mathbf{r}') = \sum_{ab} \gamma_{ab} \phi_a(\mathbf{r}) \phi_b(\mathbf{r}'),  
\end{equation}
normalized here to $N$. The method only depends on the spatial diagonal $\rho_0(\mathbf{r}) = \gamma(\mathbf{r}, \mathbf{r})$.
The 2-RDM is written as $\Gamma_{ab,cd}$, again spin-summed, corresponding to 
\begin{equation}
\Gamma(\mathbf{r}_1, \mathbf{r}_{2}; \mathbf{r}_1', \mathbf{r}_2') = \sum_{abcd} \Gamma_{ab,cd} \phi_a(\mathbf{r}_1) \phi_b(\mathbf{r}_1') \phi_c(\mathbf{r}_2) \phi_d(\mathbf{r}_2'),
\end{equation}
with normalization $N(N-1)$. The method only depends on the spatial diagonal (pair density), $P_0(\mathbf{r}_1, \mathbf{r}_2) = \Gamma(\mathbf{r}_1, \mathbf{r}_2; \mathbf{r}_1, \mathbf{r}_2)$.

To compute the matrix elements of sec~\ref{sec:theory} we need the 1-RDM and 2-RDM of the monomers and the integrals of the functions $b_i(\rv)$ with the spatial orbitals, which, with the choice of eq~\eqref{eq:bi}, are all of the kind
\begin{equation}
I_{stu}^{ab} = \int (x-x_0)^s (y-y_0)^t (z-z_0)^u \phi_a(\mathbf{r}) \phi_b(\mathbf{r}) \mathrm{d} \mathbf{r}
\end{equation}
For every monomer we need to calculate $S_{ij}$ of eq~\eqref{eq:Sij}, $\tau_{ij}$ of eq~\eqref{eq:tauij} and $\mathbf{d}_i$ of eq~\eqref{eq:di} from the 1-RDM, and $P_{ij}$ of eq~\eqref{eq:Pij} and $\mathbf{D}_i$ of eq~\eqref{eq:Di} from the 2-RDM. We first write all the matrix elements by assuming that the constraint of eqs~\eqref{eq:densconstrA}-\eqref{eq:densconstrB} is satisifed, which amounts to assuming
\begin{equation}\label{eq:pi}
p_i = \frac{1}{N}\int   b_i(\mathbf{r}) \rho(\mathbf{r}) \mathrm{d} \mathbf{r} = \sum_{ab} \frac{\gamma_{ab}}{N}I^{ab}_{s_i,t_i,u_i}=0.
\end{equation}
When this does not hold, we make the appropriate modifications in terms of $p_i$, see eqs~\eqref{eq:Sijcons}-\eqref{eq:Dicons} below.

We then have for the matrix $S_{ij}$ of eq~\eqref{eq:Sij}
\begin{equation}
S_{ij} = \sum_{ab} \gamma_{ab} I^{ab}_{s_i+s_j,t_i+t_j,u_i+u_j},
\end{equation}
and for $\tau_{ij}$ of eq~\eqref{eq:tauij}
\begin{equation}
\begin{split}
\tau_{ij} = s_i s_j \sum_{ab} \gamma_{ab} I^{ab}_{s_i+s_j-2, t_i+t_j, u_i + u_j} + t_i t_j \sum_{ab} \gamma_{ab} I^{ab}_{s_i+s_j, t_i+t_j-2, u_i + u_j} \\
+ u_i u_j \sum_{ab} \gamma_{ab} I^{ab}_{s_i+s_j, t_i+t_j, u_i + u_j-2}
\end{split}
\end{equation}
The components of the vector $\mathbf{d}_i$ of eq~\eqref{eq:di} are given by the dipole moment in directions $e=x,y,z$. For example for the $x$-direction:
\begin{equation}
d_{x,i}   = \sum_{ab} \gamma_{ab} I^{ab}_{s_i+1, t_i, u_i},
\end{equation}
while for $y$ and $z$ we get analogous expressions with $I^{ab}_{s_i, t_i+1, u_i}$ and $I^{ab}_{s_i, t_i, u_i+1}$, respectively. For convenience, we also define (with analogous expressions for the $y$ and $z$ directions),
\begin{equation}\label{eq:dx0}
d_{x, 0} = \int (x-x_0) \rho(\mathbf{r}) \mathrm{d} \mathbf{r} = \sum_{ab} \gamma_{ab}  I^{ab}_{1,0,0}.
\end{equation}
The matrix $P_{ij}$ of eq~\eqref{eq:Pij}, which is a sort of overlap mediated by the pair density, is given by
\begin{equation}
P_{ij} = \sum_{abcd} \Gamma_{ab,cd} I^{ab}_{s_i, t_i, u_i} I^{cd}_{s_j, t_j, u_j}.
\end{equation}
For the components of the vector $\mathbf{D}_i$ of eq~\eqref{eq:Di} we have, for example in the $x$ direction,
\begin{equation}
D_{x, i} = \sum_{abcd} \Gamma_{ab,cd} I^{ab}_{1, 0, 0} I^{cd}_{s_i, t_i, u_i},
\end{equation}
with similar expressions with $I^{ab}_{0, 1, 0}$ and $I^{ab}_{0, 0, 1}$ for the other two components.
When $p_i$ of eq~\eqref{eq:pi} is not zero we need to modify the matrix elements according to
\begin{align}
S_{ij} &= \sum_{ab} \gamma_{ab} I^{ab}_{s_i+s_j,t_i+t_j,u_i+u_j} - N p_i p_j\label{eq:Sijcons}\\
d_{x,i} &= \sum_{ab} \gamma_{ab} I^{ab}_{s_i+1, t_i, u_i} - p_i d_{x, 0}\\
P_{ij} &= \sum_{abcd} \Gamma_{ab,cd} I^{ab}_{s_i, t_i, u_i} I^{cd}_{s_j, t_j, u_j} - p_i p_j N (N-1)\\
D_{x, i} &= \sum_{abcd} \Gamma_{ab,cd} I^{ab}_{1, 0, 0} I^{cd}_{s_i, t_i, u_i} - (N-1) p_i d_{x, 0}, \label{eq:Dicons}
\end{align}
with analogous expressions for the components $y$ and $z$ of $\mathbf{d}_i$ and $\mathbf{D}_i$, and $d_{x,0}$ defined in eq~\eqref{eq:dx0}.
\subsection{Implementation}
The expression for the dispersion coeffcients of eqs~\eqref{eq:C6ani} and \eqref{eq:C6iso}, with the computational details just described, has been written in Python and interfaced with PySCF\cite{Sun-WIRCMS-18} and HORTON\cite{Horton2}. The Python package is open-source and available on Github (\url{https://github.com/DerkKooi/fdm}). The reduced density matrices of the monomers are obtained from PySCF and the multipole moment integrals are calculated using HORTON. 
The monomer densities and pair densities have been computed at three different levels of theory: Hartree-Fock, MP2 and CCSD, where for open-shell systems we used Restricted Open-Shell Hartree-Fock (ROHF).
The geometries of the molecules were optimised using the ORCA program package\cite{Nee-WIRCMS-12} using MP2 level of theory with def2-TZVPPD basis set.

\subsection{Choice of the basis set for the monomer calculations}
\begin{figure}[!ht]
\begin{center}
\includegraphics[width=0.32\textwidth]{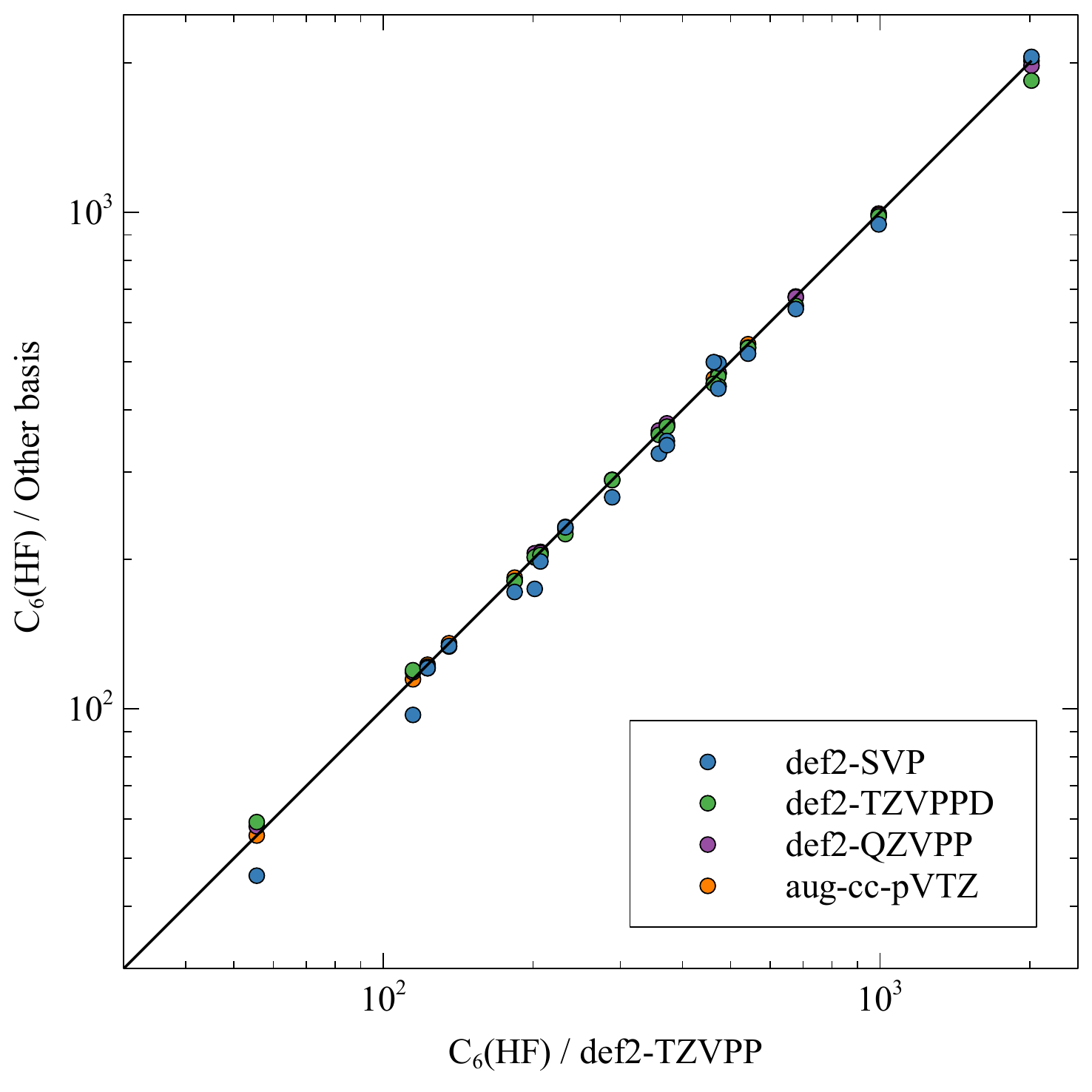}\hfill
\includegraphics[width=0.32\textwidth]{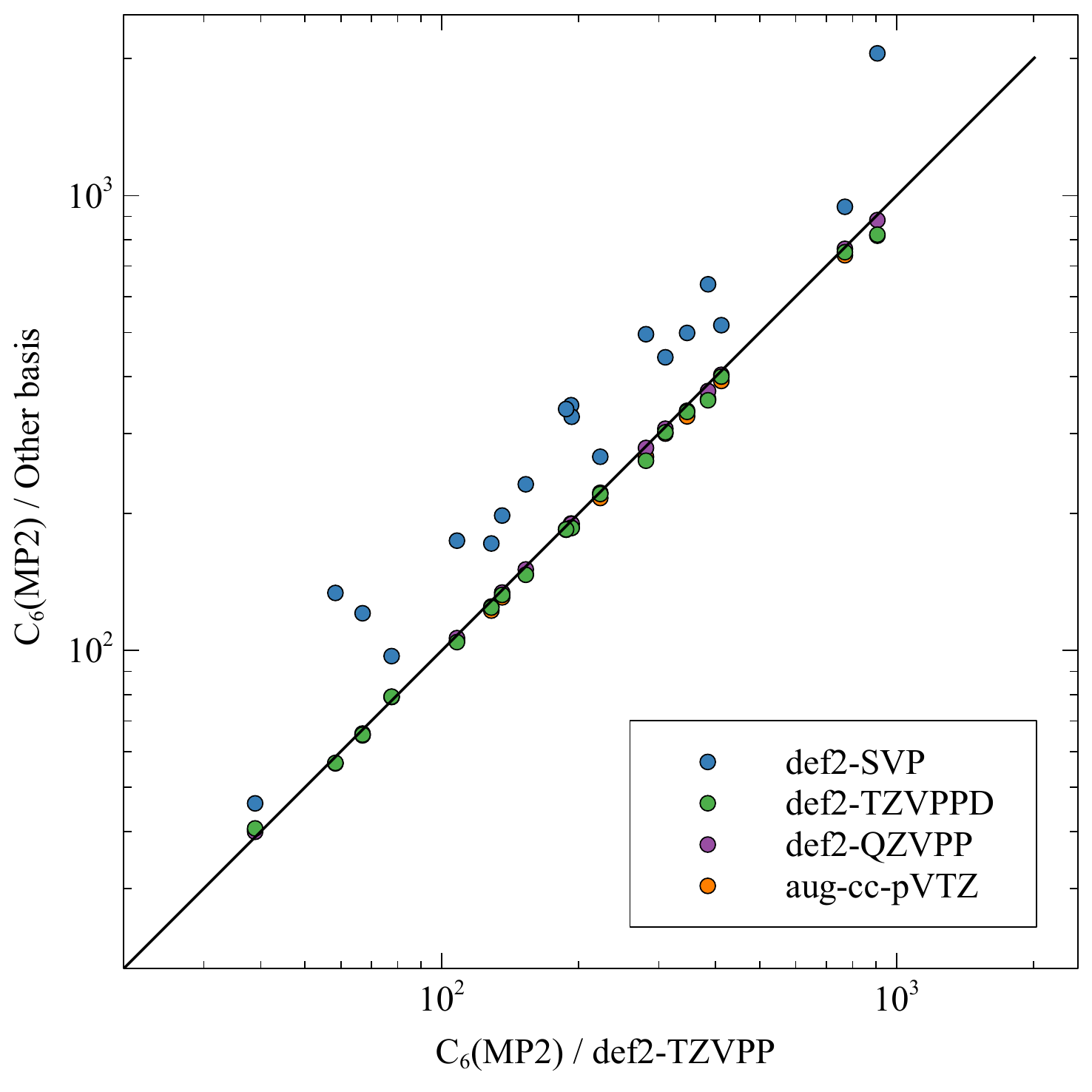}\hfill
\includegraphics[width=0.32\textwidth]{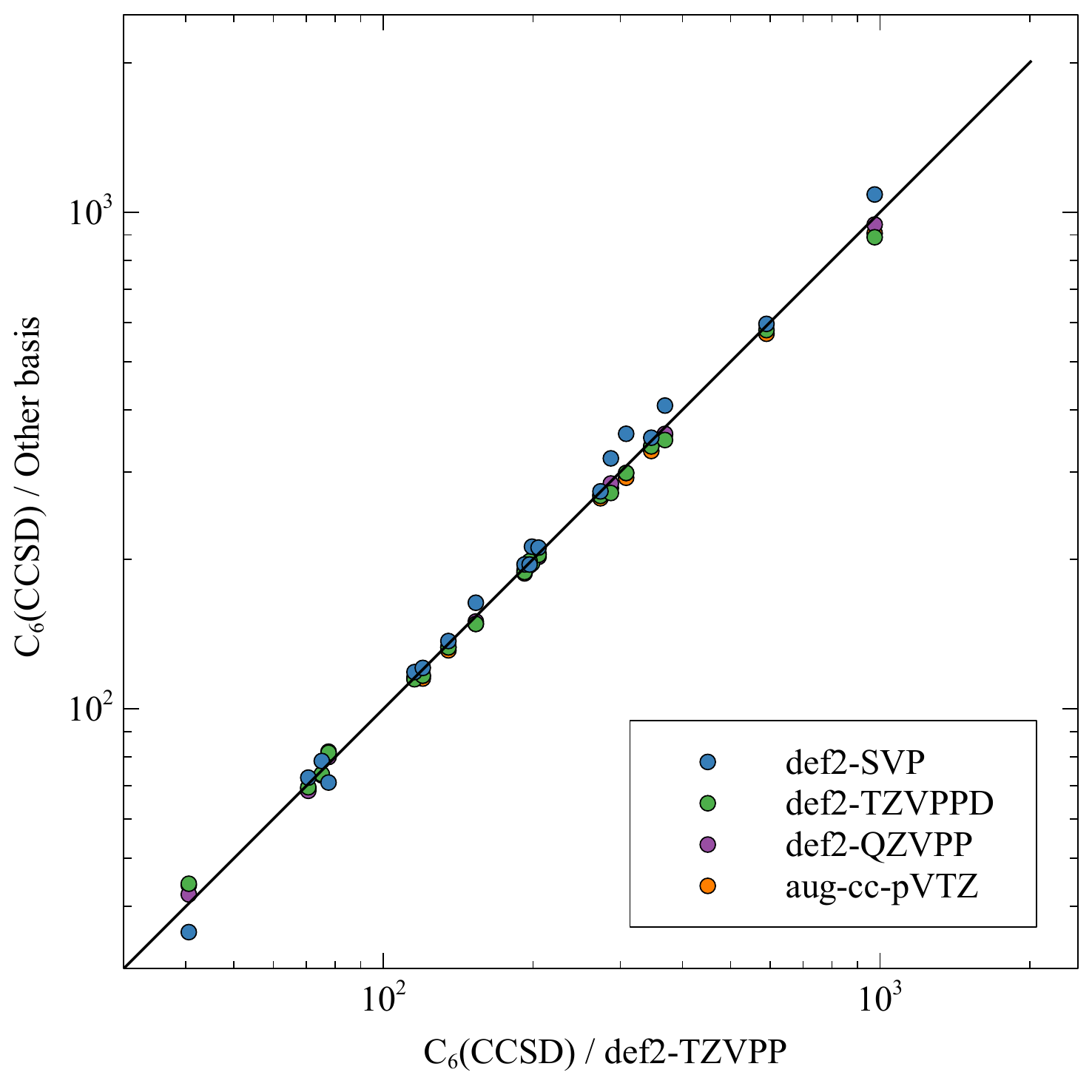}
\end{center}
\caption{Isotropic $\overline{C}^{AA}_6$ dispersion coefficients for molecules calculated using Hartree--Fock, MP2 and CCSD pair densities with different basis sets compared with calculations done using def2-TZVPP basis set. Coefficients calculated with the MP2 pair density are the most sensitive to the basis set used, while Hartree--Fock and CCSD methods produce quite robust results. The largest outlier for all the methods used is the \ce{CS2} molecule, which is further discussed in section \ref{subsec:convergence_n} and in fig~\ref{fig:molecules-multipole-convergence}.}
\label{fig:basis-set-comparison}
\end{figure}

We have extensively explored the dependence on the basis set used for the monomer pair densities calculations for all but the largest molecules, finding that, in general, going beyond a def2-TZVPP (or equivalent) quality does not particularly improve the overall results, with few singular exceptions. The mean absolute percentage errors (MAPE) for dispersion coefficients of molecules obtained with def2-QZVPP basis set differs from the def2-TZVPP ones from 1.5 to 2.2\%, with Hartree--Fock being the least and CCSD the most sensitive. When diffuse functions are incorporated into the basis set, the MAPE difference between def2-TZVPP and def2-TZVPPD basis sets range from 2.6 to 3.5\%, with Hartree--Fock being the least sensitive and MP2 the most sensitive. These differences are less than half the MAPE with respect to the reference values.
As a representative example, in fig~\ref{fig:basis-set-comparison} we show the $\overline{C}_6^{AA}$ for the molecules considered here with HF, MP2 and CCSD pair densities using different basis sets compared with calculations done using the def2-TZVPP basis set, which is our choice for all the results presented in the next section~\ref{sec:results}. 

We should also remark that, since Hartree-Fock pair densities usually lead to an overestimation of $C_6$, if one uses a smaller double-$\zeta$ basis set the performance in this case usually improves as the smaller basis makes the overestimation less profound. The results obtained using a correlated pair density, however, become worse if we go below triple-$\zeta$ quality. 

All our results obtained with different basis sets are available in the supplementaty material.


\subsection{Convergence with respect to the number $n_{\rm max}$ of $b_i$ functions}
\label{subsec:convergence_n}

\begin{figure}[!ht]
\begin{center}
\includegraphics[width=0.42\textwidth]{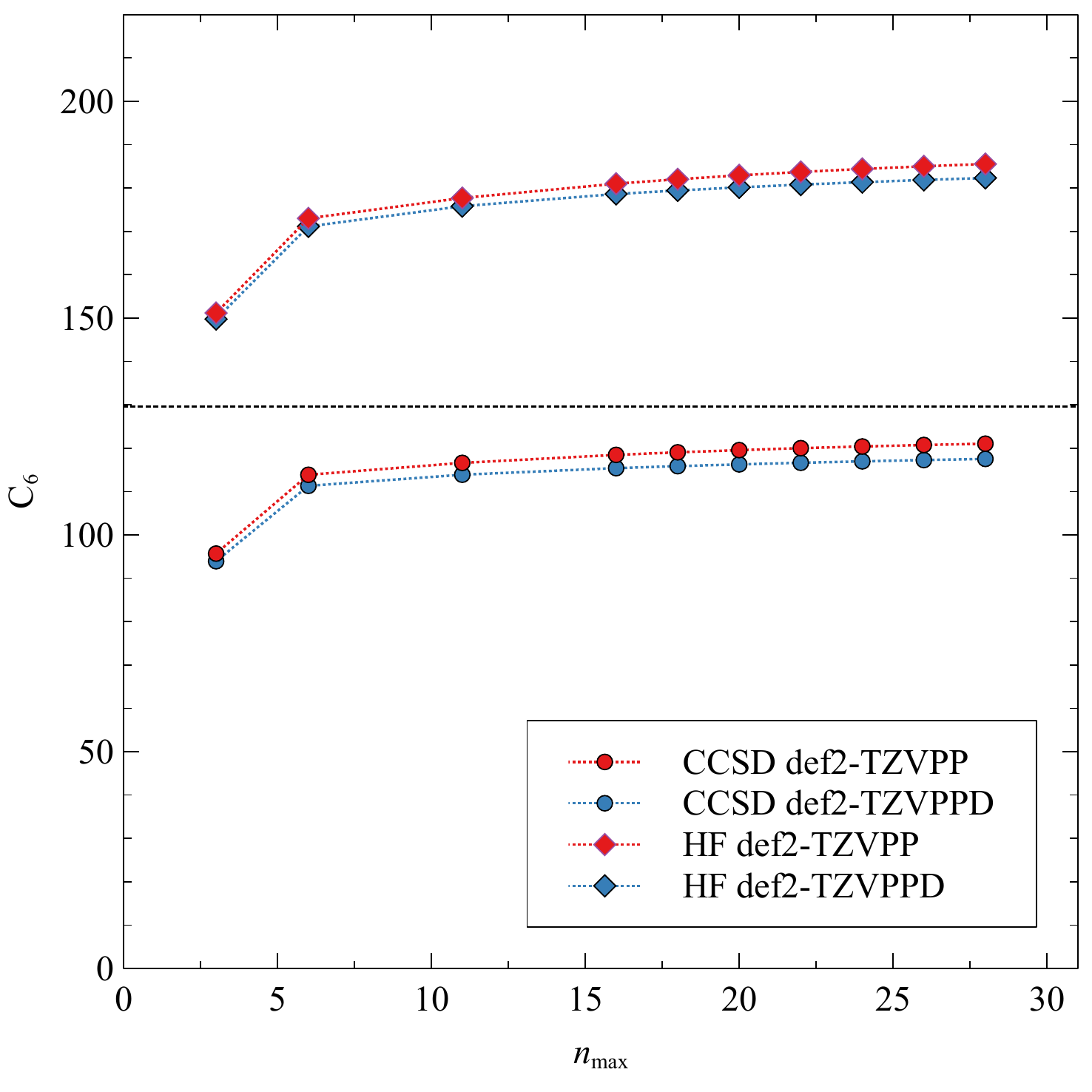}\hfill
\includegraphics[width=0.42\textwidth]{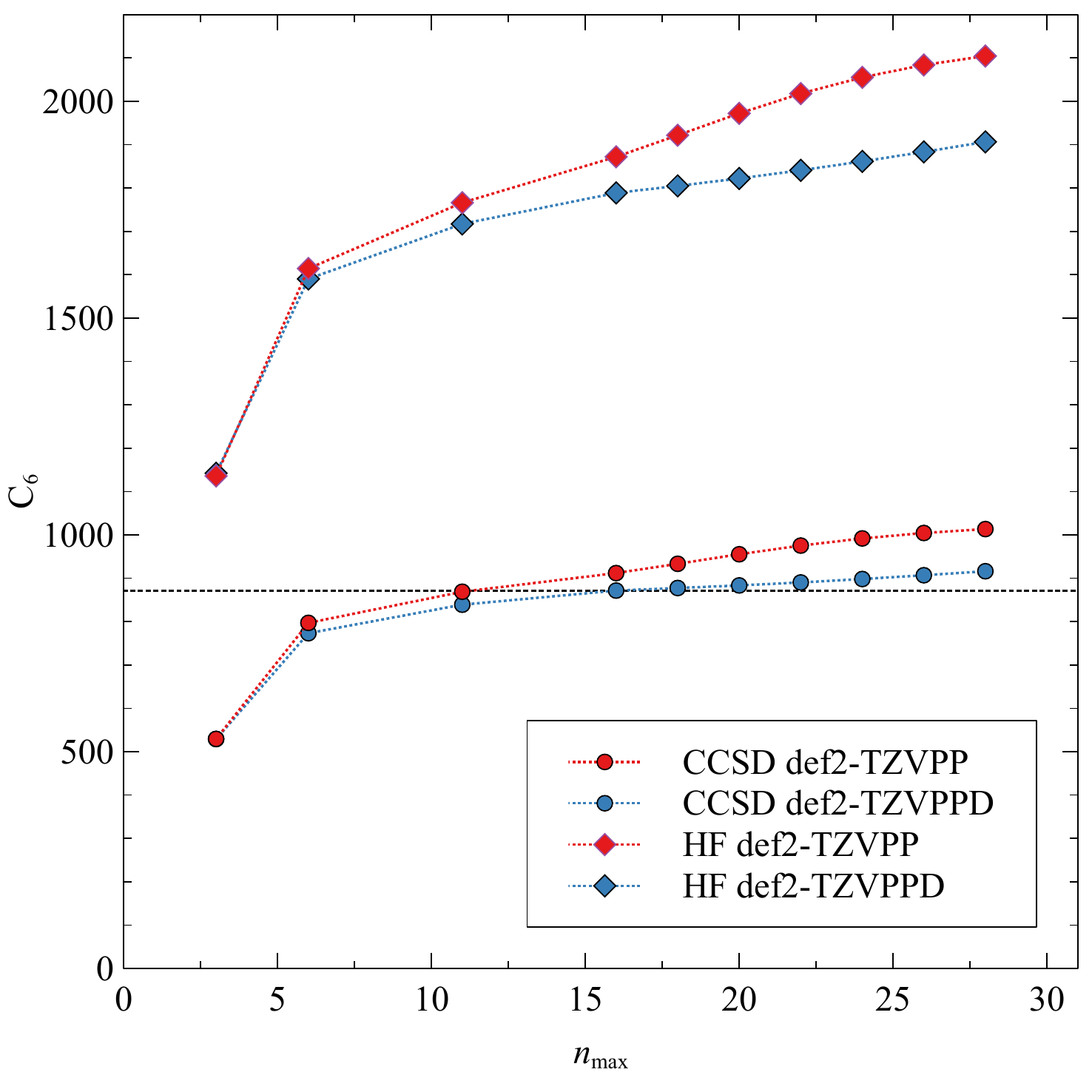}
\end{center}
\caption{Isotropic $\overline{C}^{AA}_6$ coefficients for \ce{CH4} (left) and  \ce{CS2} (right) as a function of $n_{\rm max}$, related to the number of functions used to expand the function $J_R(\rv, \rv')$ of eq~\eqref{eq:Jwithb}, for Hartree-Fock and CCSD pair densities of the monomers in different basis sets. Reference values from DOSD measurements are shown with a dotted line. The case of \ce{CH4} is representative for what we have observed for the vast majority of systems. \ce{CS2} is the worst case found: it is clearly not well converged and needs diffuse functions in the basis set. }
\label{fig:molecules-multipole-convergence}
\end{figure}

In all our calculations we have fixed $n_{\rm max}=22$, which yields in general well converged results for the vast majority of cases, and it is also a value for which the multipole integrals are numerically stable. However, we should remark that there are a few cases in which the convergence with the number of $b_i$ functions has not been satisfactorily reached. 
As a typical example for how the vast majority of systems behave, we show in the left panel of  fig~\ref{fig:molecules-multipole-convergence}, the convergence of $\overline{C}_6^{AA}$ for \ce{CH4} with respect to $n_{\rm max}$, for both HF and CCSD pair densities, with and without diffuse functions in the basis set for the monomer calculation. We see that the result is well converged and that the addition of diffuse functions has little effect, with CCSD underestimating the $C_6$ coefficient.
There are however three molecules (\ce{SO2}, \ce{CS2} and \ce{CO2}) where the values between $n_{\rm max}=20$ and $n_{\rm max}=22$ deviate more than 1\%. The worst case is \ce{CS2}, shown in the right panel of the same fig~\ref{fig:molecules-multipole-convergence}: we see that even at $n_{\rm max}=28$ the dispersion coefficient of \ce{CS2} is not converged and that CCSD overestimates $C_6$ . The inclusion of diffuse functions, in this case, improves both the convergence profile and the accuracy.

\section{Results}\label{sec:results}
Dispersion coefficients were computed for five data sets:
\begin{enumerate}
\item $C_6^{AA}$ for 23 atoms and ions,
\item $C_6^{AB}$ for 253 mixed pairs consisting of atoms and ions, 
\item isotropic $\overline{C}_6^{AA}$ for a set of 26 molecules,
\item isotropic $\overline{C}_6^{AB}$  for a set of 157 mixed molecule pairs,
\item anisotropic $\Gamma_6^{AB}$ and $\Delta_6^{AB}$ (where applicable) for three diatomics and interacting with noble-gas atoms. 
 \end{enumerate}
 In all cases we compare our results with reference values obtained from dipole oscillator strength distribution (DOSD) data computed\cite{JiaMitCheBro-ATDNDT-15} or constructed from measurements and theoretical constraints.\cite{ZeiMea-MP-77,JhaMea-MP-80,JhaMea-CJC-84,KumMea-CP-84,KumMea-MP-85,PazKumThuMea-CJC-88,KumMea-MP-92,KumKumMea-MP-02,Kum-JMS-02,KumKumMea-MP-03,KumKumMea-CP-03,KumJhaMea-CCCC-05,KumJhaMea-CJC-07,KumMea-MP-08,MeaKum-IJQC-90,KumMea-CP-94}

\subsection{Dispersion coefficients for atoms and ions}

\begin{figure}
\begin{center}
\includegraphics[width=0.45\textwidth]{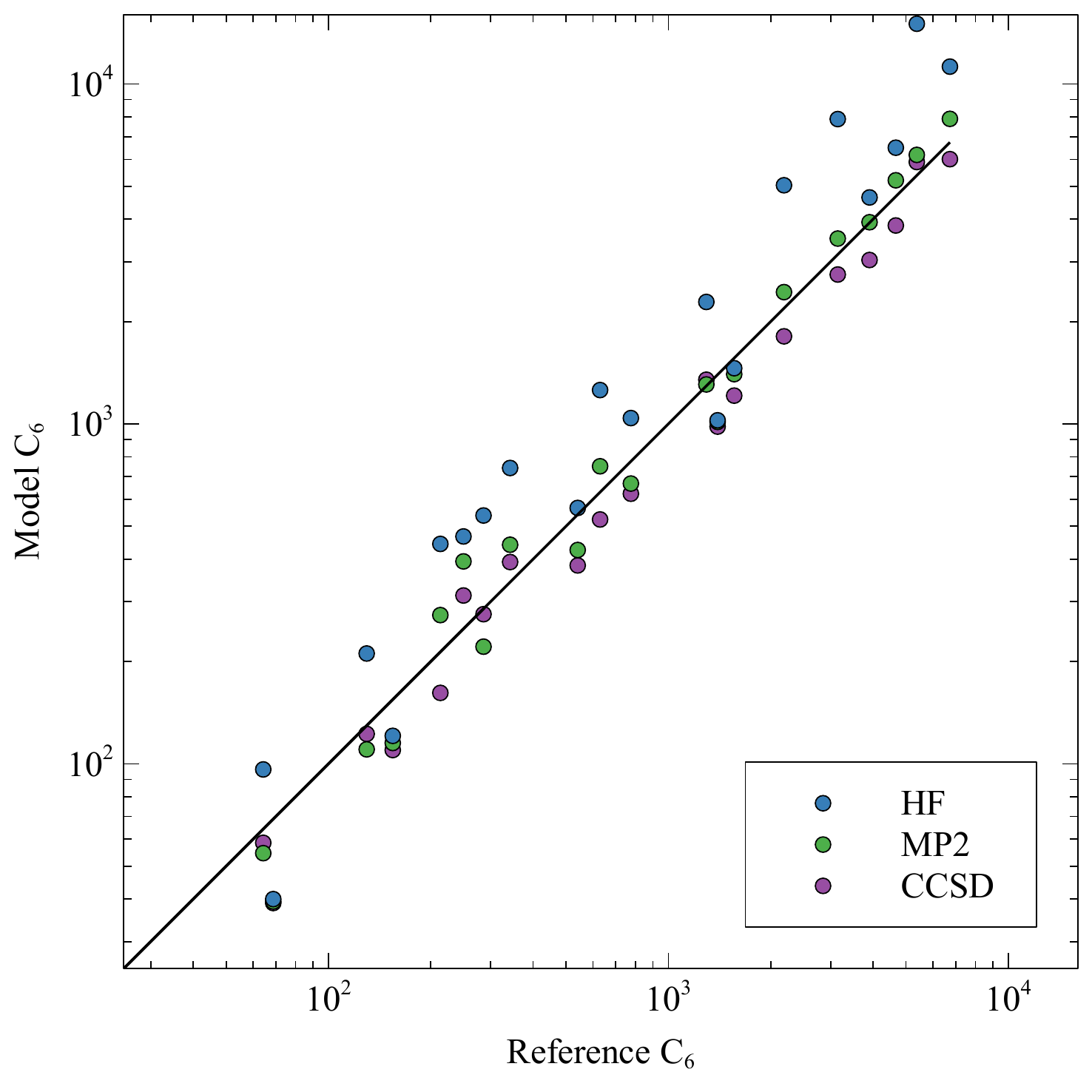}
\end{center}
\caption{Results for $C_6^{AA}$ for 23 atoms and ions (see table~\ref{table:atomic-coeff}). The solid line depicts one-to-one correspondence of the model with the reference data obtained from ref~\citenum{JiaMitCheBro-ATDNDT-15}. The mean absolute percentage error (MAPE) for Hartree-Fock, MP2 and CCSD monomer pair densities is 62.1\%, 17.7\% and 16.2\%, respectively.}\label{fig:atomic}
\end{figure}

The results for set 1, using HF, MP2 and CCSD pair densities (def2-TZVPP basis set, with effective core potential (ECP) for fifth and sixth row elements) for the monomers are presented in table \ref{table:atomic-coeff} and compared with accurate reference data.\cite{JiaMitCheBro-ATDNDT-15} The MAPE for Hartree-Fock, MP2 and CCSD monomer pair densities is 62.1\%, 17.7\% and 16.2\%, respectively. The same results are also illustrated in fig~\ref{fig:atomic}. 
Notice that the result for H in table \ref{table:atomic-coeff} has a small residual error of 1.2\% due to the basis set used, since  the results for $C_6^{AA}$ (as well as $C_8^{AA}$ and $C_{10}^{AA}$) from our wavefunction are exact when the exact hydrogenic orbital is used.\cite{KooGor-JPCL-19}

\begin{figure}
\begin{center}
\includegraphics[width=0.45\textwidth]{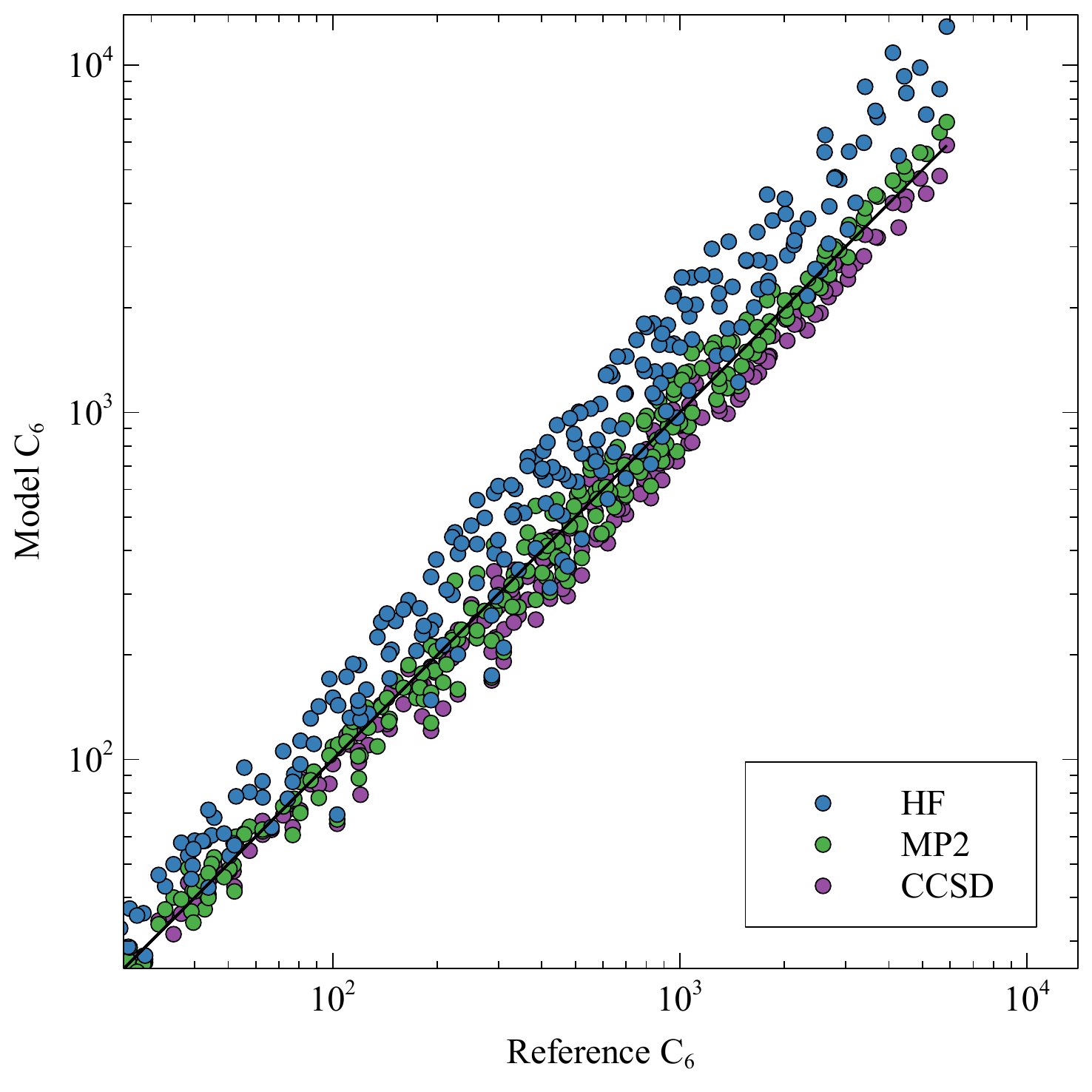}
\end{center}
\caption{Dispersion coefficients $C_6^{AB}$ for 253 pairs formed by selecting $A$ and $B$ from the species listed in table \ref{table:atomic-coeff}, computed using Hartree--Fock, MP2 and CCSD pair densities for the monomers with the def2-TZVPP basis set. The solid line depicts one-to-one correspondence of the model with the reference data obtained from ref~\citenum{JiaMitCheBro-ATDNDT-15}. The mean absolute percentage error (MAPE) for Hartree-Fock, MP2 and CCSD monomer pair densities is 52.3\%, 12.1\%, 11.9\%, respectively. }
\label{fig:atomic-mixed}
\end{figure}

For the test set 2, the different pairs are formed by selecting $A$ and $B$ from the species listed in table \ref{table:atomic-coeff}. The results for the dispersion coefficients $C^{AB}_6$ computed using different pair densities for the monomers, again with the def2-TZVPP basis set, are compared to accurate reference values\cite{JiaMitCheBro-ATDNDT-15} in fig~\ref{fig:atomic-mixed}.  The MAPE for Hartree--Fock, MP2 and CCSD are slightly better, being 52.3\%, 12.1\%, 11.9\%, respectively. All the values obtained are available in the supplementary material. 

These results for atoms and ions are not extremely promising, in particular because they do not always improve with the accuracy of the theory used to treat the monomers. From fig~\ref{fig:atomic-mixed}, it is evident that the use of the Hartree-Fock pair densities leads to an overestimation of the dispersion coefficients. However, for some systems (Li, Na, \ce{Be+}, \ce{Mg+}) the FDM method combined with correlated pair densities considerably underestimates the dispersion coefficient,  and in those cases Hartree--Fock pair densities yield better results than MP2 and CCSD ones. As it should, the CCSD pair density tends to produce a lower bound for the dispersion coefficient, but with some exceptions (e.g. Ag, Cu, \ce{Ba+}). 

The picture improves considerably if we look at closed-shell species only: if we consider the 15 noble-gas pairs, the MAPE for HF, MP2 and CCSD pair densities is 41.9\%, 12.2\% and 4.3\%, respectively. 
Also, if we consider the subset of our dataset formed by the 45 pairs of the noble gas and alkali elements used by Becke and Johnson\cite{BecJoh-JCP-05} in their original paper on the exchange-hole dipole moment (XDM) dispersion model (see their table I), we obtain MAPE for MP2 and CCSD equal to 9.6\% and 7.7\%, respectively, lower than the one of XDM (11.4\%), while with Hartree-Fock pair densities our MAPE is 27.3\%. 

Overall, these first results indicate that the constrained $\method$ ansatz can work well for closed-shell species, while being less reliable for open shell cases. As we shall see in the next sec~\ref{subsec:C6molecules}, the results for the isotropic dispersion coefficients for closed-shell molecules are reasonably accurate and robust, confirming these first findings.

\begin{table}
\caption{Dispersion coefficients $C_6^{AA}$ for a set of atoms and ions computed using Hartree--Fock, MP2 and CCSD pair densities for the monomers with the def2-TZVPP basis set, with effective core potential (ECP) for fifth and sixth row elements. For each species, the dispersion coefficient closest to the reference value is in bold font. The mean absolute percentage error (MAPE) as well as the maximum absolute percent deviation (AMAX) for the data set are reported. }
\label{table:atomic-coeff}
\begin{tabular}{|l|r|r|r|r|}
\hline
 {Species} & {Ref.\cite{JiaMitCheBro-ATDNDT-15}}&  {HF} & {MP2} & {CCSD} \\ 
\hline
H	&	6.50	&	{\bf 6.42}	&	{\bf 6.42}	&	{\bf 6.42}  \\
Li	&	1395.80	&	{\bf 1024.59}	&	1013.58	&	981.77 \\
Na	&	1561.60	&	{\bf 1458.17}	&	1400.47	&	1211.02 \\
K	&	3906.30	&	4636.05	&	{\bf 3919.56}	&	3034.83 \\
Rb	&	4666.90	&	6493.38	&	{\bf 5207.21}	&	3833.89 \\
Cs	&	6732.80	&	11244.92	&	7894.26	&	{\bf 6008.81} \\
Cu	&	249.56	&	466.54	&	393.98	&	{\bf 312.73}\\
Ag	&	342.29	&	741.72	&	441.06	&	{\bf 392.27}\\
\ce{Be+}	&	68.80	&	{\bf 40.00}	&	39.36	&	38.95 \\
\ce{Mg+}	&	154.59	&	{\bf 120.87}	&	115.17	&	109.78 \\
\ce{Ca+}	&	541.03	&	{\bf 565.94}	&	425.62	&	383.40 \\
\ce{Sr+}	&	775.72	&	1040.23	&	{\bf 667.53}	&	623.11 \\
\ce{Ba+}	&	1293.20	&	2284.40	&	{\bf 1306.28}	&	1348.81\\
Be	&	213.41	&	443.51	&	273.87	&	{\bf 161.69} \\
Mg	&	629.59	&	1257.52	&	750.44	&	{\bf 523.40} \\
Ca	&	2188.20	&	5035.02	&	{\bf 2441.13}	&	1809.39 \\
Sr	&	3149.30	&	7882.73	&	{\bf 3508.83}	&	2750.55 \\
Ba	&	5379.60	&	15037.42	&	6184.94	&	{\bf 5892.73} \\
He	&	1.46	&	1.62	&	{\bf 1.43}	&	{\bf 1.43} \\
Ne	&	6.38	&	6.79	&	5.91	&	{\bf 6.19} \\
Ar	&	64.30	&	96.28	&	54.60	&	{\bf 58.57} \\
Kr	&	129.56	&	211.12	&	110.30	&	{\bf 122.45} \\
Xe	&	285.87	&	537.65	&	221.15	&	{\bf 275.55} \\
\hline
  & MAPE    & 62.1	\% &  17.7\%   & 16.2\% \\
  & AMAX    & 179.5	\% &  57.9\%   & 43.4\% \\
\hline	
\end{tabular}
\end{table}

\subsection{Isotropic dispersion coefficients for molecules}\label{subsec:C6molecules}

\begin{figure}[!ht]
\begin{center}
\includegraphics[width=0.45\textwidth]{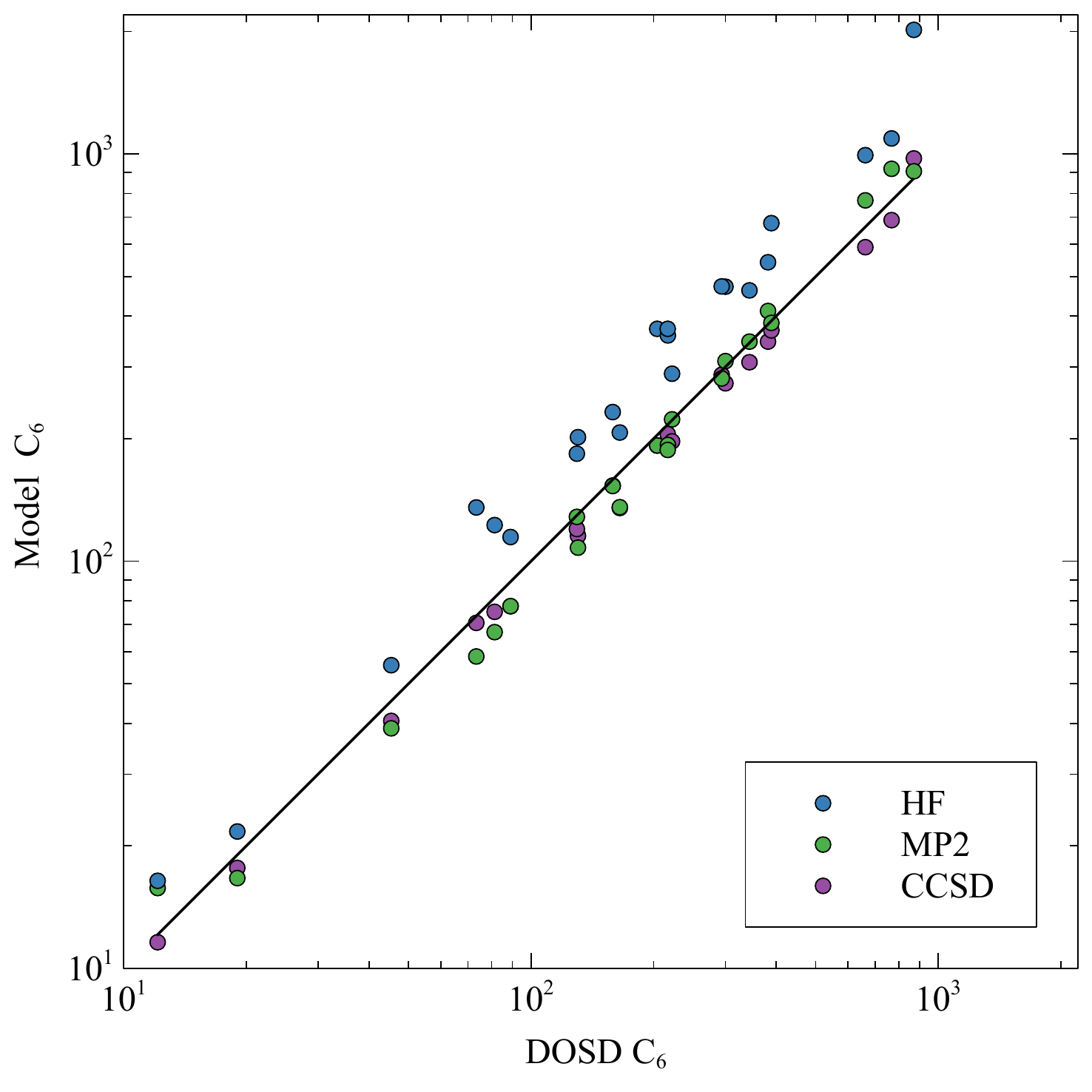}
\end{center}
\caption{The isotropic dispersion coefficients $\overline{C}_6^{AA}$ for molecules calculated using Hartree--Fock, MP2 and CCSD pair-densities and reported in table~\ref{table:molec-coeff}. The solid line depicts one-to-one correspondence of the model with the reference values.  }
\label{fig:molecules}
\end{figure}

Isotropic molecular dispersion coefficients $\overline{C}_6^{AA}$ were computed for 26 molecules consisting mainly of first and second row elements. Our results are compared with reference values calculated from DOSD\cite{ZeiMea-MP-77,JhaMea-MP-80,JhaMea-CJC-84,KumMea-CP-84,KumMea-MP-85,PazKumThuMea-CJC-88,KumMea-MP-92,KumKumMea-MP-02,Kum-JMS-02,KumKumMea-MP-03,KumKumMea-CP-03,KumJhaMea-CCCC-05,KumJhaMea-CJC-07,KumMea-MP-08} in table \ref{table:molec-coeff}, and are also illustrated in figure \ref{fig:molecules}. 
The MAPE using Hartree--Fock pair-density with def2-TZVPP basis is 52.5\%. This comes down to 13.7\% and 8.6\% when using MP2 and CCSD pair-densities, respectively. This is in line with the results for the noble-gas atoms: there is now a clear systematic improvement with the level of theory of the monomer pair densities, with CCSD yielding good results with the lowest variance.

\begin{table}
\caption{Isotropic dispersion coefficients $\overline{C}_6^{AA}$ for a set of molecules calculated using def2-TZVPP basis set. For each species, the dispersion coefficient closest to the reference value is in bold font.}
\label{table:molec-coeff}
\begin{tabular}{|l|r|r|r|r|}
\hline
 \text{Species} & \text{Ref.} & \text{HF} 
  & \text{MP2} 
  & \text{CCSD} 
   \\
\hline
\ce{H2}	&	12.1	\cite{ZeiMea-MP-77}	&	16.42	&	15.76	&	{\bf 11.60}	\\
\ce{C2H6}	&	381.9	\cite{JhaMea-MP-80}	&	542.19	&	{\bf 411.68}	&	346.17	\\
\ce{C2H4}	&	300.2	\cite{KumJhaMea-CJC-07}	&	472.25	&	{\bf 310.09}	&	273.50	\\
\ce{C2H2}	&	204.1	\cite{KumMea-MP-92}	&	372.13	&	{\bf 192.44}	&	192.34	\\
\ce{H2O}	&	45.3	\cite{ZeiMea-MP-77}	&	55.57	&	38.88	&	{\bf 40.55}	\\
\ce{H2S}	&	216.8	\cite{PazKumThuMea-CJC-88}	&	358.57	&	193.01	&	{\bf 199.08}	\\
\ce{NH3} 	&	89	\cite{ZeiMea-MP-77}	&	114.66	&	{\bf 77.58}	&	77.52	\\
\ce{SO2}	&	293.9	\cite{KumMea-CP-84}	&	473.00	&	281.11	&	{\bf 287.00}	\\
\ce{SiH4}	&	343.9	\cite{KumKumMea-CP-03}	&	462.54	&	{\bf 346.18}	&	308.09	\\
\ce{N2}	&	73.3	\cite{ZeiMea-MP-77}	&	135.53	&	58.37	&	{\bf 70.57}	\\
\ce{HF}	&	19	\cite{KumMea-MP-85}	&	21.70	&	16.67	&	{\bf 17.66}	\\
\ce{HCl}	&	130.4	\cite{KumMea-MP-85}	&	201.66	&	108.00	&	{\bf 115.47}	\\
\ce{HBr}	&	216.6	\cite{KumMea-MP-85}	&	372.17	&	187.65	&	{\bf 205.25}	\\
\ce{H2CO}	&	165.2	\cite{KumMea-JCMSE-02}	&	207.04	&	{\bf 135.75}	&	135.13	\\
\ce{CH4}	&	129.6	\cite{JhaMea-MP-80}	&	183.70	&	{\bf 128.52}	&	120.00	\\
\ce{CH3OH}	&	222	\cite{KumJhaMea-CCCC-05}	&	288.75	&	{\bf 223.07}	&	196.93	\\
\ce{CS2}	&	871.1	\cite{KumMea-CP-84}	&	2017.72	&	{\bf 906.90}	&	975.32	\\
\ce{CO}	&	81.4	\cite{JhaMea-CP-82}	&	122.69	&	66.98	&	{\bf 75.13}	\\
\ce{CO2}	&	158.7	\cite{JhaMea-CP-82}	&	232.44	&	153.00	&	{\bf 153.45}	\\
\ce{Cl2}	&	389.2	\cite{KumKumMea-MP-02}	&	676.48	&	{\bf 384.86}	&	368.85	\\
\ce{C3H6}	&	662.1	\cite{KumJhaMea-CJC-07}	&	993.49	&	769.43	&	{\bf 590.50}	\\
\ce{C3H8}	&	768.1	\cite{JhaMea-MP-80}	&	1092.09	&	919.72	&	{\bf 688.07} \\
\ce{C4H8} & 1130.2 \cite{KumJhaMea-CJC-07} & 1699.13 & 1527.79 & {\bf 1023.72}\\
\ce{C4H10} & 1268.2 \cite{JhaMea-MP-80} & 1821.45 & 1714.29 & {\bf 1137.31}\\
\ce{C5H12} & 1905.0 \cite{JhaMea-MP-80} & 2733.69 & 2872.35 & {\bf 1695.39}\\
\ce{C6H6} & 1722.7 \cite{KumMea-MP-92} & 3116.90 & 2148.87 & {\bf 1630.94}\\
\hline
 &	MAPE & 52.1\% & 13.7\% & 8.6\% \\
 &	AMAX & 131.6\% & 50.8\% & 18.2\% \\
 \hline
\end{tabular}
\end{table}

\begin{figure}[!ht]
\begin{center}
\includegraphics[width=0.45\textwidth]{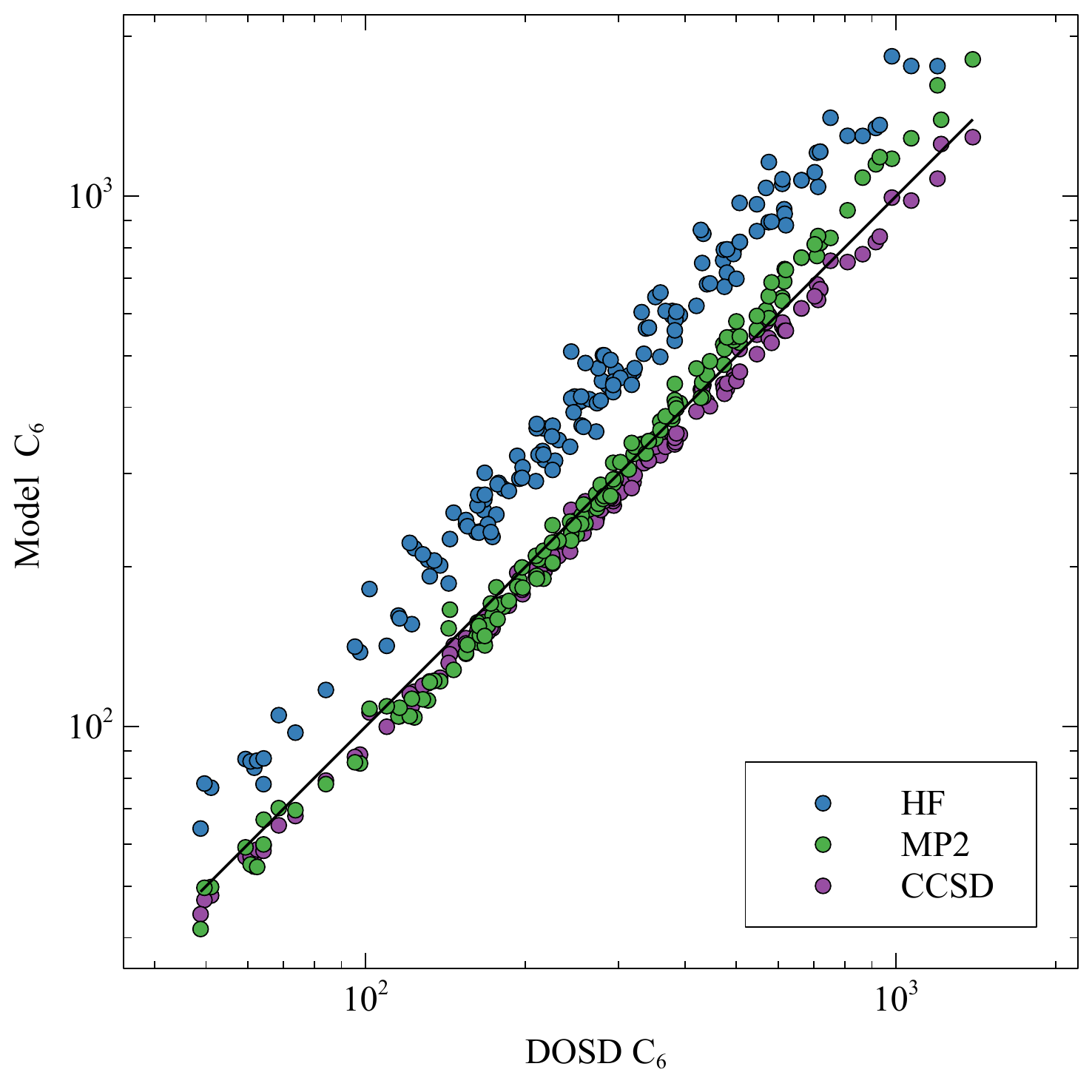}
\end{center}
\caption{Isotropic dispersion coefficients $\overline{C}_6^{AB}$ for 157 mixed molecule pairs selected from table~\ref{table:molec-coeff} calculated using Hartree--Fock (MAPE  57.1\%), MP2 (MAPE 7.9\%) and CCSD (MAPE 7.2\%) monomer pair-densities  (def2-TZVPP basis set). The solid line depicts one-to-one correspondence of the model with the reference values.}
\label{fig:mixed-pairs}
\end{figure}

For test set 4, we have computed isotropic dispersion coefficients $\overline{C}^{AB}_6$ for 157 mixed molecule pairs selected from table~\ref{table:molec-coeff}. The results are illustrated in figure \ref{fig:mixed-pairs} where they are compared, again, with reference values from DOSD, measurements\cite{ZeiMea-MP-77,JhaMea-MP-80,JhaMea-CJC-84,KumMea-CP-84,KumMea-MP-85,PazKumThuMea-CJC-88,KumMea-MP-92,KumKumMea-MP-02,Kum-JMS-02,KumKumMea-MP-03,KumKumMea-CP-03,KumJhaMea-CCCC-05,KumJhaMea-CJC-07,KumMea-MP-08} and are available in the supplementary material. The MAPE using Hartree--Fock, MP2 and CCSD pair densities are 57.1\%, 7.9\% and 7.2\%, respectively.

From these calculations we can confirm that for closed-shell molecules, the Hartree--Fock pair density leads to consistent overestimation of the dispersion coefficients. The use of a correlated pair density (MP2 or CCSD) improves the results considerably, with CCSD providing better accuracy and lowest scattering of the results. For CCSD, 11 of the 26 $\overline{C}_6^{AA}$ deviate from the reference value more than 10\%, compared to 16 for MP2 and 26 for Hartree--Fock.  With the exception of \ce{H2CO}, all the CCSD values are within 13\% of the reference value. 

Regarding the FDM results with Hartree-Fock pair densities, we should stress that in this work we are testing the formalism as derived by our trial FDM wavefunction when we assume that an exact description of the monomers is used.\cite{KooGor-JPCL-19,KooGor-FD-20} However, if we use Hartree-Fock wavefunctions for the monomers we could also revise the formalism to take into account that, in Hartree-Fock theory, part of the intramonomer electron-electron interaction is described in terms of the off-diagonal elements of the one-body reduced density matrix (1RDM), which do change in the FDM, and are quadratic in the variational parameters, like the kinetic energy. This and other flavour of approximations using Kohn-Sham orbitals and DFT xc holes will be extensively tested in future works.

\subsection{Anisotropic dispersion coefficients}
To test the applicability of the method for the orientation dependence (anisotropy) of the dispersion coefficients, we performed calculations for the diatomics \ce{H_2}, \ce{N_2} and \ce_{CO}, and interacting with noble-gas atoms. The resulting $\Gamma_6^{AB}$ are listed in table \ref{table:molec-coeff-gamma} and the resulting $\Delta_6^{AB}$ are listed in table \ref{table:molec-coeff-delta}. For both $\Gamma_6^{AB}$ and $\Delta_6^{AB}$ CCSD (MAPE 6.9\%, 7.4\%, respectively) performs better than MP2 (MAPE 40.3\%, 58.9\%), which in turn performs better than Hartree-Fock (MAPE 111.1\%, 210.2\%). MP2 performs better for the pairs involving \ce{H_2} than for the pairs involving \ce{N_2} and \ce{CO}.

\begin{table}
\caption{Anisotropic dispersion coefficients $\Gamma_6^{AB}$ for the diatomics \ce{H_2}, \ce{N_2} and \ce_{CO}, and interacting with noble-gas atoms calculated using def2-TZVPP basis set. For each species, the coefficient closest to the reference value is in bold font. The MAPE and AMAX are calculated for the product $C_6 \Gamma_6^{AB}$.}
\label{table:molec-coeff-gamma}
\begin{tabular}{|l|r|r|r|r|}
\hline
 \text{Pairs} & \text{Ref.} & \text{HF} 
  & \text{MP2} 
  & \text{CCSD} 
   \\
\hline
\ce{H2-H2} & 0.1006\cite{MeaKum-IJQC-90} & 0.1416 & 0.1099  & \textbf{0.1021}\\
\ce{H_2-N2}	 & 0.1109\cite{MeaKum-IJQC-90} &  0.1350 & \textbf{0.1040} & 0.0972\\
\ce{N_2-H_2} & 0.0966\cite{MeaKum-IJQC-90} &  0.1884 & 0.0474 & \textbf{0.1251}\\
\ce{N_2-N_2} & 0.1068\cite{MeaKum-IJQC-90} &  0.1809 & 0.0442 & \textbf{0.1211}\\
\ce{H_2-He}	 & 0.0924\cite{MeaKum-IJQC-90} & 0.1288 & 0.1013 & \textbf{0.0947}\\
\ce{H_2-Ne}	 & 0.0901\cite{MeaKum-IJQC-90} & 0.1240 & 0.0981 & \textbf{0.0920}\\
\ce{H_2-Ar}	 & 0.0971\cite{MeaKum-IJQC-90} & 0.1343 & 0.1046 & \textbf{0.0977}\\
\ce{H_2-Kr}	& 0.0986\cite{MeaKum-IJQC-90} & 0.1369 & 0.1059 & \textbf{0.0990}\\
\ce{H_2-Xe}	 & 0.1005\cite{MeaKum-IJQC-90} &0.1397 & 0.1078 & \textbf{0.1006}\\
\ce{N_2-He}	 & 0.1027\cite{MeaKum-IJQC-90} & 0.1738 & 0.0429 & \textbf{0.1192}\\
\ce{N_2-Ne}	 & 0.0999\cite{MeaKum-IJQC-90} & 0.1672 & 0.0412 & \textbf{0.1164}\\
\ce{N_2-Ar}	 & 0.1074\cite{MeaKum-IJQC-90} & 0.1800 & 0.0446 & \textbf{0.1214}\\
\ce{N_2-Kr}	&	0.1087\cite{MeaKum-IJQC-90} & 0.1827 & 0.0452 & \textbf{0.1223}\\
\ce{N_2-Xe}	 & 0.1104\cite{MeaKum-IJQC-90} & 0.1856 & 0.0461 & \textbf{0.1234}\\
\ce{CO-CO} & 0.094\cite{KumMea-CP-94} &  0.1013 & 0.0600 & \textbf{0.0956}\\
\ce{CO-H_2} & 0.0949\cite{KumMea-CP-94} & 0.1030 & 0.0616 & \textbf{0.0970}\\
\ce{H_2-CO} & 0.0976\cite{KumMea-CP-94} & 0.1350 & 0.1047 & \textbf{0.0979}\\
\ce{CO-N_2} & 0.0939\cite{KumMea-CP-94} & 0.1014 & 0.0598 & \textbf{0.0954}\\
\ce{N_2-CO} & 0.1077\cite{KumMea-CP-94} &  0.1808 & 0.0446 & \textbf{0.1216}\\
\ce{CO-He} & 0.093\cite{KumMea-CP-94} & 0.0997 & 0.0591 & \textbf{0.0947}\\
\ce{CO-Ne} & 0.0916\cite{KumMea-CP-94} & 0.0975 & 0.0580 & \textbf{0.0933}\\
\ce{CO-Ar} & 0.0942\cite{KumMea-CP-94} & 0.0975 & 0.0600 & \textbf{0.0955}\\
\ce{CO-Kr} & 0.0943\cite{KumMea-CP-94} & 0.1016 & 0.0603 & \textbf{0.0958}\\
\ce{CO-Xe} & 0.0944\cite{KumMea-CP-94} & 0.1021 & 0.0608 & \textbf{0.0961}\\
\hline
 &	MAPE & 111.1\% & 40.3\% & 6.9\% \\
 &	AMAX & 213.7\% & 67.3\% & 23.1\% \\
 \hline
\end{tabular}
\end{table}

\begin{table}
\caption{Anisotropic dispersion coefficients $\Delta_6^{AB}$ for the diatomics \ce{H_2}, \ce{N_2} and \ce_{CO} calculated using def2-TZVPP basis set. For each species, the coefficient closest to the reference value is in bold font. The MAPE and AMAX are calculated for the product $C_6 \Delta_6^{AB}$.}
\label{table:molec-coeff-delta}
\begin{tabular}{|l|r|r|r|r|}
\hline
 \text{Pairs} & \text{Ref.} & \text{HF} 
  & \text{MP2} 
  & \text{CCSD} 
   \\
\hline
\ce{H2-H2} & 0.0108\cite{MeaKum-IJQC-90} & 0.0214 & 0.0128 &\textbf{0.0110}\\
\ce{H_2-N2}	 & 0.0114\cite{MeaKum-IJQC-90} & 0.0269 & 0.0053 & \textbf{0.0126}\\
\ce{N_2-N_2} & 0.0121\cite{MeaKum-IJQC-90} & 0.0346 & 0.0021 & \textbf{0.0151}\\
\ce{CO-CO} & 0.0090\cite{KumMea-CP-94} & 0.0104 & 0.0037 & \textbf{0.0092}\\
\ce{CO-H_2} & 0.0094\cite{KumMea-CP-94} & 0.0142 & 0.0066 & \textbf{0.0096}\\
\ce{CO-N_2} & 0.0103\cite{KumMea-CP-94} & 0.0188 & 0.0028 & \textbf{0.0118}\\
\hline
 &	MAPE & 210.2\% & 58.9\% & 7.4\% \\
 &	AMAX & 427.0\% & 85.9\% & 19.8\% \\
 \hline
\end{tabular}
\end{table}

\section{Conclusions and Perspectives}\label{sec:conc}
 The ``fixed-diagonal matrices'' (FDM) idea\cite{KooGor-JPCL-19,KooGor-FD-20} provides a framework to build new approximations for the dispersion energy in terms of the ground-state pair densities (or the exchange-correlation holes) of the monomers, without the need of polarizabilities. The underlying supramolecular wavefunction describes a simplified physical mechanism for dispersion, in which only the kinetic energy of the monomers can change. While this is not what happens in the exact case, where all the terms in the isolated monomer hamiltonian change with respect to their ground-state values, for one-electron fragments the FDM still provides the exact second-order Rayleigh-Schr\"odinger dispersion energy.\cite{KooGor-JPCL-19,KooGor-FD-20} The purpose of this work was to investigate how accurate the FDM description can be for systems beyond the simple H and He cases when using a good pair-density of the monomers, focusing on the $C_6$ dispersion coefficients. In the present implementation the computational cost of the step needed on top of the monomers' ground-state calculation is $\mathcal{O}(N^4)$.
 
\begin{figure}[!ht]
\begin{center}
\includegraphics[width=0.45\textwidth]{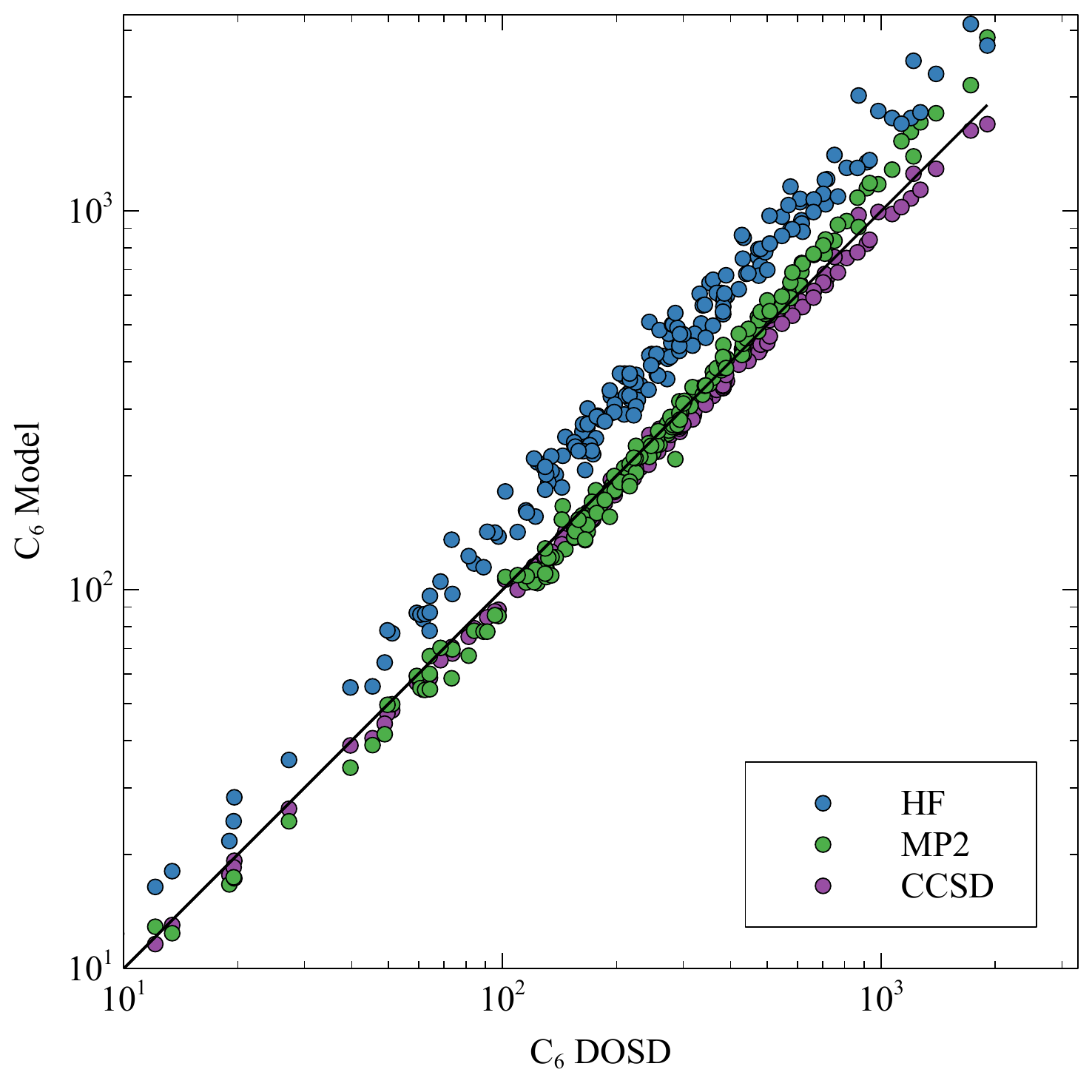}
\end{center}
\caption{Isotropic dispersion coefficients for the whole closed-shell data sets (all molecules and noble-gas atoms) calculated using Hartree--Fock (MAPE  55.3\%, AMAX 131.6\%), MP2 (MAPE 8.9\%, 50.78\%) and CCSD (MAPE 7.1\%, AMAX 18.2\%) monomer pair-densities  (def2-TZVPP basis set). The solid line depicts one-to-one correspondence of our FDM method with the reference values.}
\label{fig:allclosedshell}
\end{figure}

We have found that for closed shell species FDM yields rather accurate isotropic dispersion coefficients when using CCSD (or even MP2) monomer pair densities, with mean absolute percentage errors (MAPE) for CCSD for the whole closed-shell data set (all noble gas atoms and molecule pairs, summarized in fig~\ref{fig:allclosedshell})  of 7.1\% and a maximum absolute error (AMAX) within 18.2\%. 
FDM on top of CCSD ground states also predicts the anisotropy of dispersion coefficients, which on a limited set of pairs involving diatomics and noble-gas atoms yields satisfactory results for the anisotropy $\Gamma_6^{AB}$ (MAPE 6.9\%, AMAX 23.1\%) and the anisotropy $\Delta_6^{AB}$ (MAPE 7.4\%, AMAX 19.8\%).

From this study it also emerged that the basis set used in the monomer calculations has little effect on the computed dispersion coefficients, as the results are essentially converged at the triple-$\zeta$ level. 
Although not competitive with linear-response (LR) CCSD based methods at the complete basis set (CBS) limit, which can achieve accuracy\cite{KorPrzJez-MP-06,VisBuc-JCP-19} of 1-3\%, FDM combined with CCSD pair densities seems to have similar accuracy (for closed-shell systems) of LR-CCSD-based methods when triple- or double-zeta basis sets are used for the latter.\cite{KorPrzJez-MP-06,Gob-THESIS-16} 
From the data available, the FDM dispersion coefficients calculated using CCSD pair densities also outperform XDM for both atoms\cite{BecJoh-JCP-05} and molecules.\cite{JohBec-JCP-05} The DFT-D4 dispersion model\cite{CalEhlHan-JCP-19} has notably lower MAPE for a test set consisting of closed-shell molecules i.e. test sets 3+4, but a higher AMAX (MAPE 4.3\%, AMAX 29.1\%). 
Other methods like TS\cite{TkaSch-PRL-09} and LRD\cite{SatNak-JCP-09} have also a similar or slightly better performance than FDM with CCSD. The TD-DFT results based on  Hartree-Fock derived response functions for all atoms and ions reported by Gould and Bucko\cite{GouBuc-JCTC-16} can achieve accuracy between 1 and 5\%, and the more refined MCLF method of Manz et al.\cite{ManCheCol-RSCA-19} has again a MAPE of 4.5\% for a set of closed-shell molecules.
An advantage over methods like DFT-D4 is that FDM also predicts the anisotropy of dispersion coefficients, and gives access not only to energetics but also to a wavefunction that can be used in various frameworks. 
We should also remark that the performance of our method is less satisfactory for open-shell atoms and ions.

The main motivation for this work is to provide a solid basis for constructing DFT approximations based on a microscopic real-space mechanism for dispersion, given by a simple competition between kinetic energy and inter-monomer interaction. Before making approximations for the exchange-correlation holes of the monomers it was important to assess how accurate the method can be when good pair densities are used. 
Considering that the FDM is parameter-free and does not use the polarizabilities as input, the results for closed-shell systems are satisfactory, indicating that the simplified physical mechanism behind it, although not exact, is a reasonable approximation. In our view, this is also conceptually interesting, as it indicates that it is possible to describe reasonably well the overall raise in energy of the monomers with kinetic-energy-only effects.

In future works we will investigate possible ways to improve the results for open shell fragments, and we will work on building approximations based on model exchange-correlation holes from density functional theory, but also revisiting the formalism in the Hartree-Fock framework by taking into account the effects of the change in the off-diagonal elements of the 1RDM on the monomer's Fock operators.
We should also stress that the choice of the basis in which to expand the density constraint, i.e. the dispersal functions $b_i(\rv)$ of eq~\eqref{eq:Jwithb}, is arbitrary and that the current choice of eq~\eqref{eq:bi} is far from optimal. For some cases, the convergence with the number of dispersal functions is slow, and too high multipole moments integrals may become numerically unstable. We will thus also explore more closely the determination of an optimal choice for the dispersal functions, as we have preliminary indications\cite{KooGor-FD-20} that with a proper choice it is possible to use just a few of them to obtain well converged results.

\begin{acknowledgement}
D.P.K. and P.G.-G. acknowledge financial support from the Netherlands Organisation for Scientific Research (NWO) under Vici Grant 724.017.001, and T.W. acknowledges financial support from the Finnish Post Doc Pool and the Jenny and Antti Wihuri Foundation. We acknowledge E. Caldeweyher and S. Grimme for providing the data to perform the comparison to D4.
\end{acknowledgement}

\begin{suppinfo}

\end{suppinfo}

\bibliography{../../references/dispersion_bib}


%
%
%
%
%

\end{document}